\documentclass[referee]{cjaa}           

\usepackage{graphicx}                   
\input{epsf.sty}                        
\input{psfig.sty}                       

\begin{document}

   \title{Muon and Tau Neutrinos Spectra from Solar Flares
}

 \author{Daniele Fargion
      \inst{}\mailto{daniele.fargion@roma1.infn.it}
      and Federica Moscato
      }

   \offprints{D. Fargion}                   

   \institute{Physics Department and Infn,  Rome University La Sapienza, P.le A. Moro,2, 00185 Rome,
   Italy\\
             \email{daniele.fargion@roma1.infn.it }
             }


 %


   \abstract{Most power-full  solar flare as the  ones
   occurred on $23th$ February $1956$, September $29th$  $1989$, $28th$
   October and  on $2nd-4th$ November  $2003$
   are sources of cosmic rays, X, gamma and neutrino bursts. These  flares took
place both on front or  in the edge and in the hidden solar disk.
 The observed and estimated total flare energy (${E}_{FL} \simeq {10}^{31}\div
{10}^{33}$ erg) should be a source   of a prompt secondary
neutrino burst originated, by proton-proton-pion production on
the   sun itself; a more delayed and spread neutrino flux signal
arise  by the  solar charged flare  particles
  reaching the terrestrial atmosphere.    These first earliest prompt solar neutrino burst might be  observed,
    in a few neutrino clustered events, in present or future largest neutrino underground
    detectors as Super-Kamiokande one, in time correlation with the X-Radio
    flare. The  onset in time correlation has great statistical significance.
    Our first estimates of neutrino signals in largest underground
    detectors hint for few events in correlation with
    X,gamma,radio onser. Our  approximated spectra for muons
    and taus from these rare solar eruption are shown over the
    most common background. The muon and tau signature is very peculiar
  and characteristic over electron and anti-electron neutrino
  fluxes. The rise of muon neutrinos will be detectable above the minimal muon threshold
  $E_{\nu}$$ \simeq 113$ MeV energy , or above the pion and $\Delta^o$ thresholds ($E_{\nu}$$ \simeq 151$ and $484$ MeV ).
    Any large  neutrino flare event record might  also verify the expected neutrino
     flavour mixing leading  to a few   as well as a comparable,
      ${\nu}_{e}$, ${\nu}_{\mu}$, $\bar{\nu}_{e}$, $\bar{\nu}_{\mu}$
        energy fluence and  spectra. The rarest tau appearence
        will be possible only for hardest solar  neutrino energies
        above $3.471$ GeV .}
\authorrunning{D. Fargion,F.Moscato}            
\titlerunning{Muon and Tau Neutrino from   solar    flare}  

\maketitle
%

\section{Introduction}           
\label{sect:intro}
The recent exceptional solar flares on October-November $2003$
remind us the rarest the historical one on  February 23th,1956,
see (F.Bachelet. A.M.Conforto $1957$; J.Simpson $1957$
\cite{ref0})and most power-full comparable event on September
$29th$, $1989$ at $11:30-12:00 U. T.$, see (Alessio, Allegri,
Fargion, Improta, Iucci, Parisi et all. $1991$; (\cite{ref1}) and
L.I.Miroshnichenko et all $2000$ (\cite{89}) ). These events were
source of high energetic charged particles whose observed
energies, $E_{p}$, range between the values: $15 GeV \geq
E_{p}\geq 100$ MeV. Even higher proton solar  energies $E_p \geq
500 GeV$ have been reported see ( Karpov et all. reference
\cite{ref9}). A large fraction of these {\itshape{primary}}
particles, i. e. solar flare cosmic rays, became a source, on the
Sun surface and later on the Earth atmospheres, of neutrons ( see
reference \cite{ref1}) as well as {\itshape{secondary}} kaons,
pions $K^{\pm}$, $\pi^{\pm}$ by their particle-particle
spallation. Their following {\itshape{secondary}} rays, as muons
$\mu^{\pm}$ and neutrinos and anti-neutrinos ${\nu}_{\mu}$,
$\bar{\nu}_{\mu}$ as well as $\gamma$ rays, and their final relic
neutrinos ${\nu}_{\mu}$, $\bar{\nu}_{\mu}$, ${\nu}_{e}$ and
relevant $\bar{\nu}_{e}$ were also released by the chain reactions
$\pi^{\pm} \rightarrow \mu^{\pm}+\nu_{\mu}(\bar{\nu}_{\mu})$,
$\pi^{0} \rightarrow 2\gamma$, $\mu^{\pm} \rightarrow
e^{\pm}+\nu_{e}(\bar{\nu}_{e})+ \nu_{\mu}(\bar{\nu}_{\mu})$.
There are two different sites for these chain decays and
consequently, two corresponding neutrino emissions (see Fargion
$1989$ , \cite{ref10}):
\begin{description}
  \item[(1) ] A first, brief and sharp solar flare {\itshape{neutrino burst}};
  \item[(2) ] A later, diluted and delayed  terrestrial {\itshape{neutrino flux}}.
\end{description}
The first is a prompt  (few seconds-minutes on-set)  neutrino
burst due to charged particles scattering in the solar shock
waves, associated with known prompt gamma-X neutron events.  The
largest event yet recorded was an $X28$ which occurred at $19:50$
UT Nov $4, 2003$. The consequent  {\itshape{solar}} flare
neutrinos reach the Earth laboratory with a well defined
direction and within a narrow time gate.

Their average energies $<E_{{\nu}_{e}}>$ are probably large (for
given proton energy respect to an event on Earth atmosphere)
because their associated primary rays (${\pi}^{\pm}$,
${\mu}^{\pm}$) decay mainly in flight at low solar densities,
i.e. where their primaries suffered negligible energy loss:
$<E_{{\nu}_{e}}>
>$ $50 MeV$, $<E_{{\nu}_{\mu}}> \simeq$ 100 $\div$ 200 MeV. The
second delayed {\itshape{neutrino flux}} originated on the
Earth's atmosphere, is due to the late arrival of prompt solar
charged particles (nearly ten minutes later than the radio-X
onset). These particles are not to be  confused with lower energy
ones in much later solar winds. These solar flare rays, even
nearly relativistic (hundreds-thousand MeV), are charged and bent
by inter-planetary particles and fields; therefore their arrival
as well as their relic neutrino emission on Earth atmosphere
takes place tens minute or even a few hours later the solar
X-radio sharp event. Therefore their signal is widely spread and
diluted in time. The terrestrial neutrino directions, at sea level
are in this terrestrial case, nearly isotropic or even more
clustered near the terrestrial magnetic poles. More over a large
fraction of their low energy primary solar flare ray (protons,
alpha) energies are dissipated by ionization in terrestrial
atmosphere. Therefore {\itshape{terrestrial}} electronic neutrinos
${\nu}_{e}$, $\bar{\nu}_{e}$, are originated by muons mostly at
rest, (because of the dense terrestrial atmosphere) and are
leading to a soft terrestrial neutrino flare spectra: their mean
energy ${E}_{{\nu}\oplus}$ is on average smaller than the original
{\itshape{solar flare}} ones: $<{E}_{{\nu}\odot}>\simeq 100$ MeV
and their total relic energy ratio (terrestrial neutrino over
solar flare): $\frac{{E}_{{\nu}\oplus}}{{E}_{{\nu}\odot}}\leq
10^{-1}$ is also poor. Because of the cross section quadratic or
linear growth with energy the terrestrial neutrino flux is harder
to be detected respect to the solar ones. Finally the terrestrial
diluted neutrino flux may be drown (out of exceptional cases and
in present detectors) within the comparable steady atmospheric
neutrino back-ground. For these reasons we may neglect, in the
following approximation, the low energetic terrestrial
{\itshape{neutrino flux}}, even if they are a certain and well
defined  source of secondaries neutrinos; statistically it will
be hard to observe their signal in present Super-Kaimokande, SK
detector, but they might be observable in new future  larger
underground detectors.  We shall consider here mainly the
observable consequences due to the first prompt solar flare
{\itshape{a solar neutrino burst}}. Moreover, we shall consider
two main proton-proton-neutrino relics  origination: those
particles scattering  outward or inward the solar surface while
pointing to the (Earth) observer, because of the very different
consequent target solar atmosphere. Most of our present estimate
of the solar flare {\itshape{neutrino burst}} is scaled on an
integral flare energy ${E}_{FL}$, which is assumed to be of the
order of ${E}_{FL}\simeq10^{31}\div10^{32}$ erg, by comparison
with the known largest solar flare event as the one in $1956$,
$1989$. The recent  solar flare spectra are unknown but their
energies are extending well above a few GeV energy necessary to
pion production. To appreciate the order of magnitude of a solar
neutrino flare on the Earth, let us compare the
$\underline{total}$ flare energy flux ${\Phi}_{FL}$, at the
Solar-Earth distance ${d}_{\odot}$ with the corresponding energy
flux observed, by the well celebrated supernov\ae~explosion
SN1987A, on February 23th, 1987, at a distance ${d}_{SN}$:
\begin{equation}\label{eqn:eqn1} \frac{{\Phi}_{FL}}{{\Phi}_{SN}} \simeq \frac{{E}_{FL}}{{E}_{SN}}(\frac{{d}_{SN}}{{d}_{\odot}})^{2}\simeq
\frac{1}{30}(\frac{E_{FL}}{10^{32}~erg})
(\frac{E_{SN}}{3\cdot10^{53}~erg})^{-1}
\end{equation}
The ratio, even being smaller than unity, it  remarks the flare
energy relevance. Naturally the SN neutrino fluence are certainly
probed both experimentally and theoretically  while the solar
flare energy  ($10^{32}~erg$) conversion in neutrino has to be probed.
However even in a more conservative frame where  only a   fraction
$\eta$ of the flare energy  is partially  converted $\eta < \sim
0.1 $ into neutrinos, the flare energy flux on the Earth is:
\begin{equation}\label{eqn:eqn2}
{\Phi}_{FL}= 3.5\cdot \eta \cdot
10^{4}~erg~cm^{-2}(\frac{{E}_{FL}}{10^{32}~erg})
\end{equation}
We know that Kamiokande  detectors observed $11$ neutrino events
from the 1987A supernov\ae~explosion. The Super-Kamiokande,
because its larger volume may observe a signal $22$ time as large.
Therefore, in this low energy $\eta < \sim 10\% $ conversion the
signal ($\geq \sim 0.8 $) is near or above unity and it may be
reached. Moreover the expected pion-muon neutrino flare mean
energy $<{E}_{{\nu}{FL }}> \geq kT_{{\nu}{SN}} \simeq 10 $ MeV is
much larger ${E}_{{\nu}{FL }}\simeq 0.1-1$ GeV, than the
corresponding one for thermal supernov\ae~ neutrinos ; the
consequent neutrino cross section with nucleons grows square with
the energies; therefore the event number in present first
approximation is  larger (by a factor ten or more) than ten MeV
energy. In conclusion, if a realistic fraction ($\simeq 0.001$) of
the total flare energy (${E}_{FL}\simeq{10}^{32}$ erg) was
emitted, by proton-pion-muon chain, into  neutrino relics, as we
shall show in more details, their arrival on Earth might be
nearly detectable ($\sim1\div5$ events) and it would be much
worth checking the Super-Kamiokande records at a the X-radio
precursor flare times corresponding, for instance to X-ray maximum
on  $28th$ October and during $2nd-4th$ November (edge), $13$
November (hidden) solar flare $ 2003$.
   Of course also gamma signals  are expected by similar event:
    there are very limited gamma data (OSSE-EGRET) on solar flare and only a very surprising evidence
    of hard gamma events  (by R. P. Lin et all. 2003,(\cite{Lin})) on recent July $2002$
    solar flare where hard X-gamma events have been followed in details by  RHESSI gamma detector;
     the gamma     energy on $2002$ flare output is  smaller  than our neutrino estimate.
      Incidentally it is worth-full to remind that  the
      absence of large gamma solar flare have been used to infer a
      bound on anti-meteorite and  anti-matter presence in our
      solar system and galaxy (see reference Fargion and Khlopov $2003$, see \cite{ref4})
    But downward gamma flare and-or  higher $X$ flare energy, beaming
   and hidden flare,  might play a key role in enhancing neutrino signal over gamma flare.
   We therefore consider the total flare energy (kinetic,X) as
    a main meter to foresee the neutrino signal.

\section{The solar flare energy}
\label{sect:the solar}
The energy released during the largest known flares is mainly in
the form of inter-planetary shock waves ${E}_{FL} \geq {10}^{32}$
erg ( see reference \cite{ref2}) up to  ${E}_{FL} \leq {10}^{33}$
erg, (see reference Lin et.all 2003 ; \cite{89}). Another large
fraction of energy is found in optical emission ${E}_{FLop}\simeq
8\cdot{10}^{31}$ erg. A considerable energy fraction is also
observed in soft  and hard X-rays (by electromagnetic or nuclear
bremsstrahlung ) as well as energetic cosmic rays
($2\div5\cdot{10}^{31}$ erg). The flares might be on Earth front
or just beyond, as the one on September $1989$ located behind the
West limb of the Sun (${105}^{o}$ West) and last event on $4th$
November $2003$; the $1989$ flare was observable
$\underline{first}$ by 8.8 Gigahertz radioburst, (because of the
refractive index of solar atmosphere), at time 11:20 U. T. and,
later it reached a higher (visible) height from where it was
observable in X-ray (see (\cite{ref1})).Is there any {\itshape
hidden} underground flare whose unique faithful trace is in a
powerful (unobserved) neutrino burst?  Gamma rays secondaries,
due to common neutral pion decay, positron annihilations and
neutron capture, have a very small cross section and there must
also be as observed a trace on the Sun surface of such a powerful
hidden flare. Nevertheless observed gamma ray flares from known
ones, are not in favor of any extreme $E_{FL} \gg 10^{33} erg$
underground flares( see \cite{ref3}). However it must be kept in
mind that the rarest event on February $'56$ was not observed in
gamma band, because of the absence of such satellite detector at
that epoch, while the $Sept. 29th 1989 $ flare was not detected
in gamma rays because it occurred on the opposite solar side.
Nevertheless lower power-full solar flare as the $4th June 1991$
have been experienced in all radio-X-gamma energy up to tens MeV
energy band also by OSSE detector in CGRO satellite. Therefore
there are no real severe direct bounds on a larger {\itshape
hidden} underground flare. One may suspect that too large flare
event should be reflected somehow into electromagnetic cascade
which may influence the continuous solar energy spectrum
(${E}_{\odot}\simeq 3.84 \cdot{10}^{33}$ erg $s^{-1}$), even in
the observable solar side. Moreover recent accurate helio
seismography  might be able to reveal any extreme hidden flare
energy. Therefore we may restrict our most powerful solar flare
energy in  range:
\begin{equation}\label{eqn:eqn3} {10}^{32}~erg \sim\geq {E}_{FL}\geq {10}^{31}~erg
\end{equation}
keeping the lowest energy  as the  flare energy threshold.
\subsection{The pion production in solar flare} \label{sect:the}
The kaon-pion-muon chain reactions and their consequent neutrino
relics spectrum in solar atmosphere may be evaluated in detail if
the energetic particle (protons, alpha nuclei, ...) energy spectra
is known, as well as the solar density and magnetic configuration.
 Indeed magnetic screening may reduce high energy particle scattering
 in the solar flare regions. Successful description for terrestrial atmospheric
 neutrinos, and their primary relic of cosmic rays, has been obtained. Our
approach, ignoring the exact spectrum for protons in recent solar
flare and the detailed magnetic configuration, will force us to
consider only averaged values, neglecting  the (higher energetic)
Kaon production. In order to find the interaction probability for
an energetic proton (${E}_{p} \simeq 2$ GeV) to scatter
un-elastically with a target proton at rest in solar atmosphere,
we must assume an exponential solar density function following
well known solar density models. (reference \cite{ref5})
\begin{equation}\label{eqn:eqn4} {n}_{\odot}={N}_{0}{e^{\frac{-h}{h_{0}}}};~
{N}_{0}=2.26\cdot{10}^{17}~{cm}^{-3},
~{h}_{0}=1.16\cdot{10}^{7}~cm
\end{equation}
where $h_{0}$ is the photosphere height where flare occurs.
\subsection{Upward protons interactions in solar flare}
\label{sect:upward}
The unelastic proton-proton cross section for energetic particles
(${E}_{p}>2$ GeV) is nearly constant:
${\sigma}_{pp}(E>2~GeV)\simeq 4\cdot{10}^{-26}~{cm}^{2}$.
Therefore the scattering probability ${P}_{up}$ for an orthogonal
up-ward energetic proton ${p}_{E}$, to produce by nuclear
reaction, pions (or kaons) is:
\begin{equation}\label{eqn:eqn5} {P}_{up}={1-{e^{-\int^{\infty}_{h_{0}=0}
{\sigma}_{pp}n_{\odot}dh}}}\simeq 0.1
\end{equation}
A terrestrial Observatory in direct line of sight with a solar
flare would observe only 10\% (or much less, if, as it is well
possible ${h}_{0}> 10^7 cm$) of the primordial proton flare
number, converted into pions and relic muons, neutrinos and
electron-positron pairs. Moreover, because of the kinematics,
only a fraction smaller than 1/2 of the energetic proton will be
released to pions (or kaons) formation. In the simplest reaction,
source of pions, ($p+p\rightarrow {{\Delta}^{++}}n\rightarrow
p{{\pi}^{+}}n$; $p+p\rightarrow
{{{\Delta}^{+}}p}^{\nearrow^{p+p+{{\pi}^{0}}}}_{\searrow_{p+n+{\pi}^{+}}}$)
at the center of mass of the resonance ${\Delta}$ (whose mass
value is ${m}_{\Delta}=1232$ MeV), the reaction ${R}_{{\pi}{p}}$
between the pion to the proton energy is:
\begin{equation}\label{eqn:eqn6} {R}_{{\pi} p}=
\frac{{E}_{\pi}}{{E}_{p}}=\frac{{{{m}_{\Delta}}^{2}}+{{{m}_{\pi}}^{2}}
-{{{m}_{p}}^{2}}}{{{{m}_{\Delta}}^{2}}+{{{m}_{p}}^{2}}-{{{m}_{\pi}}^{2}}}=0.276
\end{equation}
Therefore the total pion flare energy due to upward proton is:
\begin{equation}\label{eqn:eqn7}
{E}_{{\pi}_{FL}}=P{R}_{{\pi}p}{E}_{FL}=2.76\cdot{10}^{-2}{E}_{FL}
\end{equation}
Because of the isotopic spin the probability to form a charged
pion over a neutral one in the reactions above: $p+p\rightarrow
p+n+{\pi}^{+}$, $p+p\rightarrow p+p+{\pi}^{0}$, is given by the
Clebsh Gordon coefficients, (3/4), and by the positive-negative
ratio (1/2):
\begin{equation}\label{eqn:eqn8} {C}_{\frac{{\pi}^{-}}{{\pi}^{0}}}\simeq
{C}_{\frac{{\pi}^{+}}{{\pi}^{0}}}\simeq \frac{3}{8}
\end{equation}
The ratio of the neutrino of the muon energy in
pion decay is also a small adimensional  fraction
${R}_{{\nu}_{\mu}{\mu}}$
\begin{equation}\label{eqn:eqn9} {R}_{{\nu}_{\mu}{\mu}} =
\frac{{E}_{{\nu}_{\mu}}}{{E}_{\mu}}=\frac{{{m}_{\pi}}^{2}-{{m}_{\mu}}^{2}}
{{{m}_{\pi}}^{2}+{{m}_{\mu}}^{2}}=0.271
\end{equation}
In a first approximation one may assume that the total pion
energy is equally distributed in all its final remnants:
($\bar{\nu}_{\mu}$, ${e}^{+}$, ${\nu}_{e}$, ${\nu}_{\mu}$) or
(${\nu}_{\mu}$, ${e}^{-}$, $\bar{\nu}_{e}$, $\bar{\nu}_{\mu}$):
\begin{equation}\label{eqn:eqn10} \frac{{E}_{\bar{\nu}_{\mu}}}{2}\simeq
\frac{{E}_{{\nu}_{\mu}}}{2}\simeq {E}_{{\nu}_{e}}\simeq
{E}_{\bar{\nu}_{e}}\simeq \frac{1}{4}{E}_{{\pi}^{+}}
\end{equation}
Actually the correct averaged energy (by Michell parameters) for
neutrino decay ${\mu}^{\pm}$ at rest are:
${E}_{\bar{\nu}_{e}}={E}_{{\nu}_{e}}=\frac{3}{10}{m}_{\mu}\simeq
\frac{1}{4}{m}_{\pi};{E}_{\bar{\nu}_{\mu}}={E}_{{\nu}_{\mu}}\simeq
\frac{9}{20}{m}_{\mu}\simeq\frac{1}{3}{m}_{\pi}$. Similar
reactions (at lower probability) may also occur by proton-alfa
scattering leading to: ($p+n\rightarrow
{{\Delta}^{+}}n\rightarrow n{{\pi}^{+}}n$; $p+n\rightarrow
{{{\Delta}^{o}}p}^{\nearrow^{p+p+{{\pi}^{-}}}}_{\searrow_{p+n+{\pi}^{o}}}$).
Here we neglect their additional role due to the flavor mixing
and the dominance of previous reactions at soft flare spectra.
Therefore  ${E}_{{\nu}_{\mu}}>{E}_{{\nu}_{e}}$; however muon
neutrino from pions ${\pi}^{\pm}$ decays have a much lower mean
energy and the combined result in eq.($10$) is a good
approximation. We must consider also  the flavour mixing (in
vacuum ) that it is able to lead to an averaged neutrino energy
along the neutrino flight making a final mix-up and average
neutrino  flavour energy and flux intensity. In a first
approximation the oscillation  will lead to a $50\%$ decrease in
the muon component and it will change mainly  the  electron
neutrino component making it harder. We will keep this flavor
mixing  into account by an efficient conversion term $\eta_{\mu}
=\simeq \frac{1}{2}$  re-scaling the final muon neutrino signal
and increasing the electron spectra component.
 Because in ${\pi}$-${\mu}$ decay the ${\mu}$ neutrinos relic
are twice the electron ones, the anti-electron neutrino flare
energy is, at the birth place on Sun:
\begin{equation}\label{eqn:eqn11} {E}_{\bar{\nu}_{e}FL}\simeq
{E}_{{{\nu}_{e}}FL}  \simeq\frac{{E}_{{\nu}_{{\mu}} FL}}{2} \simeq
{P}_{up}{R}_{{\pi} p}{C}_{\frac{{\pi}^{-}}{{\pi}^{0}}}{E}_{FL}
\simeq 2.6\cdot{10}^{28}(\frac{{E}_{FL}}{{10}^{31}~erg})~erg.
\end{equation}
The corresponding neutrino flare energy and number fluxes at sea
level are:
\begin{equation}\label{eqn:eqn12}
{\Phi}_{\bar{\nu}_{e}FL} \simeq
9.15(\frac{{E}_{FL}}{{10}^{31}~erg})~erg~{cm}^{-2}
\end{equation}
\begin{equation}\label{eqn:eqn13}
{N}_{{\nu}_{e}} \simeq {N}_{\bar{\nu}_{e}} \simeq
5.7\cdot{10}^{4}(\frac{{E}_{FL}}{{10}^{31}~erg})
(\frac{<{E}_{\bar{\nu}_{e}}>}{100~MeV})^{-1} cm^{-2}
\end{equation}

This neutrino number is larger but comparable with a different
value evaluated elsewhere (reference \cite{ref6}). The flux
energy in eq.(12) is nearly $4000$ times smaller than the energy
flux in eq.($2$) and, as we shall see, it maybe nevertheless
nearly observable by present detectors. This flux at $GeV$ energy
may correspond approximately to a quarter of a day atmospheric
neutrino integral fluence (for each flavor specie ). Therefore it
may lead to just a half  of an event as on out-ward event on
$28th.$ October $2003$. The largest neutron and gamma flare
energies should be (and indeed are) comparable or even much larger
($February 1956$) than the upward neutrino flux energy in
eq.($12$). The exceptional solar flare on $Sept. 29th, 1989$ as
well as the most recent on $2nd -4th November 2003 $ took place
in the nearly hidden disk side and we may look now for their
horizontal or down-ward secondary neutrinos. Their scattering are
more effective and are offering more pion production. The
processes we describe here are analogous to the one considered
for horizontal and upward neutrino induced air-showers inside the
Earth Crust (see Fargion $2002$,reference \cite{Fargion 2002})
and nearly ultra high horizontal showers (detectable by EUSO).
The solar neutrino flare production is enhanced by a higher solar
gas density where the flare beam occurs. Moreover a beamed X-flare
may suggest  a corresponding beamed pion shower whose mild
beaming naturally increase the neutrino signal. Most of the
down-ward neutrino signal to be discussed below is generated at
mild relativistic regime as well as their pion and muon
secondaries; therefore there maybe also a mild outburst outside
with some anisotropy suppression, back toward the Earth.

\subsection{Downward, proton interactions in solar flare} \label{sect:downward}
High energetic protons flying downward (or horizontally) to the
Sun center  are crossing larger (and deeper) solar densities and
their probability of interaction ${P}_{d}$, is larger than the
previous one (${P}_{up}$). The proton energy losses due to
ionization, at the densities where most of their reaction take
place, are low in respect to the nuclear ones and most of the
proton flare energy is  converted into pion-Kaon nuclear
productions with few losses.

If the energetic proton direction is tangent or downward to the
solar core, the probability interaction is even larger than one.
Unstable and short lived pions of few GeV will decay in flight
because nuclear reaction at those solar atmosphere densities, are
small; the pion number density ${n}_{\pi}$ is described by an
evolution equation of the form:
\begin{eqnarray}
\frac{{dn}_{\pi}}{dt}&=&\int\int[{{\frac{{d}^{2}{{n}_{pE}}}{dEd{\Omega}}}{n}_{pT}{\sigma}_{pp}{v}_{pE}}-
{{\frac{{d}^{2}{{n}_{pE}}}{dEd{\Omega}}}{n}_{\pi}{\sigma}_{p{\pi}}{v}_{\pi}}-
{{\frac{{d}^{2}{{n}_{\pi}}}{dEd{\Omega}}}{n}_{\pi}{\sigma}_{{\pi}{\pi}}{v}_{\pi}]{dEd{\Omega}}}+\nonumber\\
&-&\int^{\infty}_{{m}_{\pi}}\frac{{d}^{2}{{n}_{\pi}}}{dE}{\Gamma}_{\pi}(\frac{{m}_{\pi}}{{E}_{\pi}}){dE}_{\pi};
\end{eqnarray}

where $n_{pE}$, $n_{pT}$, $n_{\pi}$ are the number density of the
flare energetic and target protons, $\sigma_{pp}(E)$,
$\sigma_{p\pi}(E)$, $\sigma_{\pi\pi}(E)$, are the p-p, p-$\pi$,
$\pi$-$\pi$ cross sections. The velocities $v_{pE}$, $v_{\pi}$ are
near the velocity of light and
$\Gamma_{\pi}=3.8\cdot10^{7}s^{-1}$. The last term in eq.$(14)$,
due to the relativistic pion decay, at solar densities as in
eq.$(4)$ and at an energy $E_{\pi}\simeq$ GeV, is nearly six order
of magnitude larger than all other terms. Therefore the pion
number density $n_{\pi}$ should never exceed the corresponding
proton number density ${n_{pE}}$; however the integral number of
all pion stable relics ($\bar{\nu}_{\mu}$, $\nu_{\mu}$, $\nu_{e}$,
$\bar{\nu}_{e}$, $e^{-}$) may exceed, in principle, the
corresponding number of proton flare, because each proton may be
a source of more than one pion chain. The proton number density,
below the photosphere ($h<0$), is described by a polytropic
solution, but it can be also approximated by a natural
extrapolation of the low in the eq.$(4)$  with a negative height
h.  It is easy to show that the interaction probability for a
relativistic proton ($E_{pE}>> $ GeV) reaches unity at depth
$h=-278$ Km which is the interaction length. At the corresponding
density ($n\odot\sim2.2\cdot10^{18}cm^{-3}$) the proton
ionization losses, between any pair of nuclear reactions are
negligible (few percent).  Unstable relic pions decay (almost)
freely are after a length
$L_{\pi}\simeq(\frac{E_{\pi}}{m_{\pi}}){\Gamma_{\pi}}^{-1}C\simeq
7.8\cdot10^{2}(\frac{E_{\pi}}{m_{\pi}})$ cm. The secondary muons
$\mu$ do not loose much of their energy ($\leq 1\%$) in
ionization, ($E_{\mu}\leq 0.1-1$ GeV) during their nearly free
decay: the muon flight distance is
$L_{\mu}=\frac{E_{\mu}}{m_{\mu}}{\Gamma_{\mu}}^{-1}C=
6.58\cdot10^{4}(\frac{E_{\mu}}{m_{\mu}})$ cm, and the ionization
losses are:
$\frac{dE_{\pi}}{dx}\simeq\frac{dE_{\mu}}{dx}\simeq10^{-5}~MeV~
cm^{-1}$. In conclusion most of the solar flare energy will
contribute to downward muon energy, with an efficiency $\eta$
near unity. At deeper regions, near $h<-700$ Km where the solar
density is ${n_{\odot}}\geq 10^{20}~cm^{-3}$, a GeV-muon will
dissipate most of its energy in ionization before decaying.
 In that case the energy ratio between muon relics and primary
protons is much smaller than unity.  Only a small fraction of
inward protons will reach by a random walk such deeper regions
and we may conclude that, in general, the flare energy relations
are:
$$E_{{\pi}FL}\equiv{\eta}E_{FL}\leq E_{FL}$$
\begin{equation}\label{eqn:eqn14} E_{\bar{\nu}_{e}FL}\simeq
E_{{\nu_{e}FL}}\simeq\frac{E_{{\nu_{\mu}FL}}}{2}\simeq
\frac{E_{\bar{\nu}_{\mu}FL}}{2} \simeq
\frac{{\eta}{C_{\frac{\pi^{-}}{\pi_{0}}}}{E_{FL}}}{4} \simeq
9.4\cdot10^{30}\eta(\frac{E_{FL}}{10^{32}~erg})~erg
\end{equation}
This result is nearly 36 times larger than the corresponding one
for {\itshape up-ward} neutrinos in eq.(10).  Terrestrial neutrino
relics from cosmic rays, related to analogous pion chain
reaction, lead to a predicted and observed asymmetry (see
T.Gaisser,T.Stanev $1989$, reference  \cite{ref7}) between
$\bar{\nu}_{e}$, $\nu_{e}$, due to the positive proton charge
predominance either in target and incident beam:
$$\frac{N_{\nu_{e}}}{N_{\bar{\nu}_{e}}}=\frac{N_{\mu^{+}}}{N_{\mu^{-}}}\simeq1.2$$  at
energies $10$ GeV $>E_{\nu}>100$ MeV. Therefore the energy
component of the observable flare should be marginally reduced in
eq.$(15)$, even assuming a low flare ($10^31 erg$) output:
\begin{equation}\label{eqn:eqn15} E_{\bar{\nu}_{e}}\simeq 7.8\cdot10^{-2}{\eta}E_{FL}=
 7.8\cdot10^{29}{\eta}(\frac{E_{FL}}{{10}^{31}~erg})~erg
\end{equation}

\section{Detectable solar flare neutrinos in SK-II}
\label{sect:detectable}
We cannot say much about the solar flare neutrino spectrum
because of our ignorance on the recent primordial proton flare
spectra. The solar flare are usually very soft. We may expect a
power spectrum with an exponent equal or larger than the cosmic
ray proton spectrum. Therefore we consider here only averaged
neutrino energy $<E_{\nu}>$ at lowest energies (below near GeV)
and we scale the result above, eq.$(16)$, for the anti-neutrino
numbers at sea level:
\begin{equation}\label{eqn:eqn16}
<N_{\bar{\nu}_{e}}> \simeq
1.72\cdot{10}^{6}{\eta}(\frac{<E_{\bar{\nu}_{e}}>}{100~MeV})^{-1}
(\frac{E_{FL}}{{10}^{31}~erg})~{cm}^{-2}
\end{equation}
\begin{equation}\label{eqn:eqn17}
<N_{\bar{\nu}_{\mu}}> \simeq
4.12\cdot10^{6}{\eta}(\frac{<E_{\bar{\nu}_{\mu}}>}{100~MeV})^{-1}(\frac{E_{FL}}{10^{31}~erg})~{cm}^{-2}
\end{equation}
We now consider the neutrino events due to these number fluxes at
Super-Kamiokande II; other detectors as SNO (and AMANDA if the
spectra was extremely hard) might also record a few events but at
much lower rate. The observable neutrino events, due to inverse
beta decay ($\bar{\nu}_{e}+p \rightarrow n+e^{+}$;
$\bar{\nu}_{\mu}+p \rightarrow \mu^{+}+n$), at Super-Kamiokande
detectors are:
\begin{equation}\label{eqn:eqn18}
N_{ev}=\sum_{i}\int\frac{dN_{\bar{\nu}_{i}}}{dE_{i}}{{\sigma}_{\bar{\nu}_{i}p}}(E_{{\nu}_{i}})N_{p_{SK}}dE_{i}
\end{equation}
$ i=e,{\mu}$.
A comparable  neutrino events, due to stimulated beta decay
(${\nu}_{e}+ n \rightarrow p+e^{-}$; ${\nu}_{\mu}+ n \rightarrow
\mu^{-}+ p $), must also take place (see an updated reference
Bodek et all. $2003$ ( see \cite{Bodek})).
 We may approximate this number with an averaged one
due to an effective neutrino energy $\bar{E}_{\nu}$:
\begin{equation}\label{eqn:eqn19}
{N}_{ev}=\sum_{i}<N_{{\bar{\nu}}_{i}}>{\sigma}_{\bar{\nu}_{i}p}(\bar{E}_{{\nu}_{i}}){N}_{p_{SK}}
\end{equation}
Where ${N}_{p_{SK}}$ is the proton number in the Super-Kamiokande
detector ${N}_{p_{SK}}=\frac{{N}_{p}}{{N}_{{H}_{2}O}}N_{nucl}$;
$N_{nucl}=22KT\cdot N_{A}=3.33\cdot10^{34}$; $ \frac{{N}_{p
}}{{N}_{{H}_{2}O}}= \frac{8}{18}$; ${N}_{p_{SK}}=7.38\cdot
10^{33}$. The cross section is an   elaborated analytical formula
(see Strumia et all. $2003$ reference \cite{strumia}) is possible.
 This expression is  in agreement with full result
within few per-mille for $ E_{\nu} \leq 300 MeV $ is
\begin{equation}
\label{naive+} \sigma(\bar\nu_e p)\approx
10^{-43}\,\mbox{cm}^2~p_e E_e ~E_\nu^{-0.07056+0.02018\ln
E_\nu-0.001953\ln^3 E_\nu}, \qquad E_e = E_\nu - \Delta
\end{equation}
 Where $\Delta = m_n - m_p$; $E_e$ is the energy of the escaping
 electron. In the simplest low-energy approximation (see Bemporad et all.,
$2002$ e.g. reference \cite{bemporad})
\begin{equation}
\label{simple} \sigma\approx 9.52\, \times 10^{-44} \frac{p_e
E_e}{\hbox{MeV}^2}\ \mbox{cm}^2,\qquad E_e=E_\nu \pm \Delta\hbox{
for }\bar{\nu}_e\hbox{ and }\nu_e,
\end{equation}
  In a more simple and direct form (see Fargion $1989$ ; see \cite{ref10}),
  at low energy ($10 MeV \leq {E}_{\bar{\nu}_{e}} \leq GeV $)
\begin{equation}\label{eqn:eqn20} {\sigma}_{\bar{\nu}_{e}p} \simeq
7.5\cdot10^{-44}(\frac{{E}_{\bar{\nu}_{e}}}{MeV})^{2}cm^{2}
\end{equation}
The expected neutrino event,  during the flare may be two fold as
we mentioned above: a solar burst and a terrestrial flux; for the
terrestrial neutrino flux (during the recent $28/29 Oct. 2003$
flare) we expect from the solar proton hitting the atmosphere and
leading to neutrinos  at least:
\begin{equation}\label{eqn:eqn21}
{N}_{ev} \cong 1.7\cdot10^{6}\cdot 7.38\cdot 10^{33}\cdot 6\cdot
10^{-40}\simeq 7.5 \cdot \eta
\end{equation}
 These events should  increase (almost doubling its flux)   the common atmospheric
neutrino flux background ($5.8$ event a day). For the prompt
neutrino solar burst in the Sun we expect (if covered or
horizontal) a similar number in a very narrow time window.
Naturally this result might be over-optimistic. In order to obtain
a more severe  and more pessimistic result we now tune our
expectation with the event number due to the well known
supernov\ae~SN1987A where we know (or we hope to know) the
primordial neutrino energy: $\sum{E}_{{\nu}_{SN}} \simeq
3\cdot10^{53}$ erg and $\bar{E}_{\bar{\nu}_{e}}\simeq 10 $ MeV. We
know (by cosmology and $Z_{0}$ width decay in LEP) that the
possible neutrino flavours states are ${N}_{F}=6$ (${\nu}_{e}$,
$\bar{\nu}_{e}$, $\nu_{\mu}$, $\bar{\nu}_{\mu}$, $\nu_{\tau}$,
$\bar{\nu}_{\tau}$). The Earth-SN1987A distance
$d_{SN}=1.5\cdot10^{23}$ cm lead to:
\begin{equation}\label{eqn:eqn22} {N_{ev}}_{\bar{\nu}_{e}}
=\frac{{N}_{{\nu}_{SN}}}{{N}_{F}}{\sigma}_{\bar{\nu}_{e}p}
(\bar{E}_{{\nu}_{SN}}){{N}_{pSK}}=
11(\frac{E_{SN}}{3\cdot10^{53}~erg})(\frac{\bar{E}_{\nu}}{10~MeV})
\end{equation}
It should be noted that the quadratic energy $\bar{E}_{\nu}$
dependence of the cross section $\sigma_{\bar{\nu}_{e}p}$ and the
inverse energy $\bar{E}_{\nu}$ relation of the neutrino flux
number leads to the linear dependence in eq.$(23)$.  However, the
inverse beta decay processes increases linearly with energy
$\bar{E}_{\nu}$ up to a value smaller than $m_{p}\sim$ GeV. Above
it the cross section in eq.(20) becomes  flat and only at higher
energies it grows linearly with the energy:
$\bar{\nu}_{e}+p\rightarrow
e^{+}+n;~~~\bar{\nu}_{\mu}+p\rightarrow \mu^{+}+n;$
${\nu}_{e}+n\rightarrow e^{-}+p;~~~{\nu}_{\mu}+n\rightarrow
\mu^{-}+n;$
\begin{equation}\label{eqn:eqn23}
\sigma_{\bar{\nu}_{e}{p}}\simeq6.2\cdot10^{-39}~cm^{2}(\frac{\bar{E}_{\bar{\nu}_{e}}}{GeV});~~~~
\sigma_{{\nu}_{e}{n}} \simeq
3.5\cdot10^{-39}~cm^{2}(\frac{\bar{E}_{\nu_{e}}}{GeV})cm^{2}
\end{equation}
The formulas above are  approximation only within an  energy
window
$E_{\nu_{\mu}},E_{\bar{\nu}_{\mu}},E_{\nu_{e}},E_{\bar{\nu}_{e}}\simeq
100-1000$ MeV.  As we shall see, we may neglect the prompt
neutrino-electron scattering processes due to charged or neutral
currents cross sections:
\begin{eqnarray}
\sigma_{{\nu}_{e}e}\simeq
9\cdot10^{-45}(\frac{\bar{E}_{\nu}}{MeV})cm^{2};~
\sigma_{{\nu}_{\mu}e}\simeq1.45\cdot10^{-45}(\frac{\bar{E}_{\nu}}{MeV})cm^{2}\nonumber\\
\sigma_{\bar{\nu}_{e}e}\simeq3.7\cdot10^{-45}(\frac{\bar{E}_{\nu}}{MeV})cm^{2};~
\sigma_{\bar{\nu}_{\mu}e}\simeq1.24\cdot10^{-45}(\frac{E_{\nu}}{MeV})cm^{2}
\end{eqnarray}
Indeed these values are nearly $100$ times smaller (at
$\bar{E}_{\nu}\sim100$ MeV) than the corresponding nuclear ones
in eq.$(23, 26)$.  We consider the neutrino flare signals at
Super-Kamiokande due to either $\bar{\nu}_{e}+p \rightarrow
n+e^{+}$ and $\bar{\nu}_{\mu}+p \rightarrow \mu^{+}+n$,
${\nu}_{e}+n\rightarrow e^{-}+p;~~~{\nu}_{\mu}+n\rightarrow
\mu^{-}+n;$
keeping in mind, for the latter, the threshold energy
  ${E}_{{\nu}_{\mu}}$,${E}_{\bar{\nu}_{\mu}}>113$ MeV. We may summarize from eq.(17)
the expectation event numbers at Super-Kamiokande as follows:
$ {N_{ev}}_{\bar{\nu}_{e}} \simeq
0.63{\eta}(\frac{\bar{E}_{\bar{\nu}_{e}}}{35
~MeV})(\frac{E_{FL}}{10^{31}~erg});~\bar{E}_{\bar{\nu}_{e}}\leq
100 ~MeV $;
$ {N_{ev}}_{\bar{\nu}_{e}} \simeq
1.58{\eta}(\frac{E_{FL}}{10^{31}~erg});~
\bar{E}_{\bar{\nu}_{e}}\geq100-1000 ~MeV $;
$ {{N}_{ev}}_{\bar{\nu}_{\mu}} \simeq
3.58{\eta}(\frac{{E}_{FL}}{{10}^{31}~erg});~
\bar{E}_{\bar{\nu}_{\mu}}\geq 200-1000 ~MeV $;
where ${\eta}\leq1$. The neutrino events in Super-Kamiokande may
be also recorded as stimulated beta decay on oxygen nuclei .
Indeed such reactions exhibit two possible channels: $\nu_{e}+O
\rightarrow F+ e^{-},~\bar{\nu}_{e}~O \rightarrow N+e^{+}$; they
have been analyzed by W. C. Haxton $1987$, see reference
(\cite{ref8}).  For this reason our preliminary estimate is just a
lower bound  for any high energetic ($E_{{\nu}_{e}}>100$ MeV)
neutrino spectrum.
\section{Conclusions: The Solar Muon and Tau  Neutrino }
 The Earth-Sun distance $D_{\oplus\odot}$ is large enough to guarantee a complete
  flavor mixing even for hundred MeV or GeV neutrino energies.
   Indeed the oscillation distance in vacuum:
   $
 L_{\nu_{\mu}-\nu_{\tau}}=2.48 \cdot10^{9} \,cm \left(
 \frac{E_{\nu}}{10^{9}\,eV} \right) \left( \frac{\Delta m_{ij}^2
 }{(10^{-2} \,eV)^2} \right)^{-1} \ll D_{\oplus\odot}=1.5\cdot
 10^{13}cm.
$
   The consequent flavor mixing will increase the average energy of the
 anti neutrino  electron component respect to its birth one.
 This will increase also the neutrino electron component while it will reduce
 the corresponding muon component leading to :
 ${\frac{\eta_{\mu}}{\eta_{e}}\simeq \frac{1}{2}}$  and  to
      ${N}_{{ev}_{\bar{\nu}_{\mu}}}\simeq {N}_{{ev}_{\bar{\nu}_{e}}}
       \simeq 2 (\frac{<{E}_{{\nu}_{\mu}}>}{200~MeV}(\frac{<{E}_{FL}>}{{10}^{31}~erg})$ ;
     ${N}_{{ev}_{{\nu}_{\mu}}}\simeq {N}_{{ev}_{{\nu}_{e}}} $
       as well as a comparable, ${\nu}_{e}$, ${\nu}_{\mu}$, $\bar{\nu}_{e}$, $\bar{\nu}_{\mu}$
        energy fluence and  spectra. The rise of muon neutrinos will be detectable above the minimal
muon threshold $E_{\nu}$$ \simeq  113$ MeV energy , or better
above the pion and $\Delta^o$ thresholds ($E_{\nu}$$ \simeq 151$
and $484$ MeV ). Its presence is unexpected by standard solar
physics. The prompt muon neutrino burst will inaugurate  a novel
solar muon neutrino astronomy.
 Any large neutrino flare event record might also verify the expected
neutrino   flavour mixing leading  to a few   as well as a
comparable   ${\nu}_{e}$, ${\nu}_{\mu}$, $\bar{\nu}_{e}$,
$\bar{\nu}_{\mu}$
        energy fluence and  spectra. The rarest tau appearance
        will be possible only for hardest solar  neutrino energies
        above $3.471$ GeV, but     this require
          a hard (${E}_{{\nu}_{\mu}} \rightarrow {E}_{{\nu}_{\tau}}$$\simeq  4 GeV$) flare spectra.
          At highest neutrino energies above $4.357$ GeV the
          $\tau$ appearance by  $\Delta^o$ resonance will be most
          favorite.

\begin{figure}[h]
 \centering
   \includegraphics[width=0.8\textwidth]{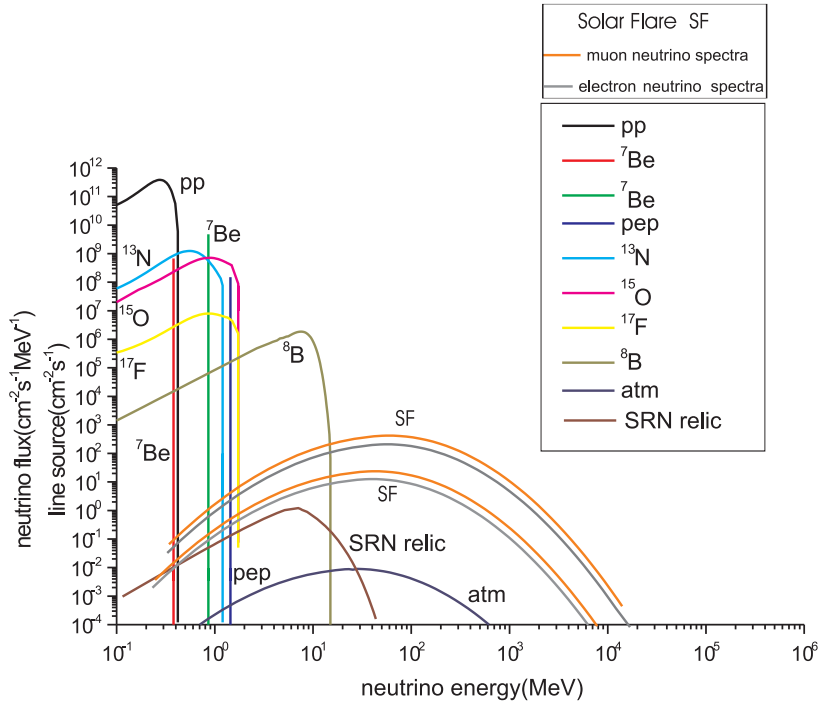}
   \caption{Solar neutrino Flare fluxes over known neutrino background: solar neutrinos by known nuclear activities
   as well as the expected Supernova Relic fluxes and the atmospheric noises.
    The Solar flare are estimated on the Earth at peak activity of the flare supposed to held $100$ $s$.
    The two primordial flavour  $\nu_e$, $\overline\nu_{e}$ and their correspondent $\nu_{\mu}$, $\overline{\nu_{\mu}}$
    are shown before their oscillation while  flying and mixing toward the Earth.
    Their final flavour are nearly equal in all states $\nu_e$,$\nu_{\mu}$, $\nu_{\tau}$
     and are not shown for sake of simplicity; however each  final $\nu_e$,$\nu_{\mu}$,
     $\nu_{\tau}$, $\overline{\nu_{e}}$,$\overline{\nu_{\mu}}$,$\overline{\nu_{\tau}}$
      fluxes are almost corresponding to the lower curve for the electron neutrino spectra.
    Both highest and lowest activities are described following an up-going (poor flux) or
    down-going  (richer flux) scenario. The primary solar flare spectra is considered like the atmosphere one
    at least within the  energy windows $E_{{\nu}_{\mu}} \simeq 10^{-3} GeV$ up to $ 10$ GeV. }
   \label{Fig:demo1}
\end{figure}

          Any positive evidence for such events will mark a new
road to Neutrino Astrophysics, to be complementary to lower
neutrino energy from Sun and Supernov\ae. New larger generations
of neutrino detectors will be more sensitive to these less
power-full, but more frequent and energetic solar flares, than
sensitive to rarest extragalactic supernov\ae~ (as the one from
Andromeda) whose time delay might give insight of neutrino mass
\cite{Fargion 2002}. In conclusion therefore the recent solar
flare on October-November $2003$ as large as the one on Sept.
$29th, 1989$ might be an exceptional source of cosmic, gamma,
neutron rays and neutrinos as well. Their minimal event number at
Super-Kamiokande ${N}_{{ev}_{\bar{\nu}_{\mu}}}\simeq
{N}_{{ev}_{\bar{\nu}_{e}}}
       \simeq 2 (\frac{<{E}_{{\nu}_{\mu}}>}{200~MeV}(\frac{<{E}_{FL}>}{{10}^{31}~erg})$ ;
     ${N}_{{ev}_{{\nu}_{\mu}}}\simeq {N}_{{ev}_{{\nu}_{e}}} $
   is near or above unity. The noise signal of energetic atmospheric neutrinos at the Japanese detector
is nearly $5.8$ event a day time corresponding to a rate
${\Gamma\simeq 6.7 10^{-5}} s^{-1}$. The minimal and the largest
predicted event number ($1\div5$)~${\eta}$, (${\eta}\leq{1}$)
within the narrow time gate defined by the sharp X burst onset
($100 s$), are above the noise.  Indeed the probability to find
by chance one neutrino event within a $1-2$ minute ${\Delta}t
\simeq 10^2 s$ in that interval is $P\simeq\Gamma
\cdot{{\Delta}T}\simeq 6.7 \cdot10^{-3}$. For a Poisson
distribution the probability to find $n=1,2, 3, 4, 5$ events in
a  narrow time window might reach extremely small values:
$ {{P}_{n}}\cong\frac{P^{n}}{n!}=( 6.7 \cdot 10^{-3}, 2.25 \cdot
10^{-5},  5 \cdot 10^{-8}, 8.39 10^{-11}, 1.1\cdot10^{-13}). $
Therefore the very possible presence of one or more high
energetic (tens-hundred MeVs) positron (or better positive muons)
 as well as any negative electron or muon,  in Super-Kamiokande
 at X-flare onset time may be a well defined and most brilliant signature of the solar
neutrino flare. A surprising discover by $\tau$ appearance  of the
complete mixing may occur by hard (${E}_{{\nu}_{\mu}} \rightarrow
{E}_{{\nu}_{\tau}}\simeq> 4 GeV$) flare spectra. Its most
recognizable  signature occurs  by a rare ${\pi}^o$ production ,
( by the common hadronic ${\tau}$ decay) , and its pion consequent
decays     into mono-cromatic $\gamma$ whose relics are consequent
$e^+$,$e^-$  relativistic pairs production. Any steep proton
flare spectrum, where a large flare energy fraction is at a low
proton energies may reduce ${\sigma}_{pp}$ un-elastic
cross-sections and increase the elastic ones, reducing the
pion-neutrino creations.  A low flare energy ${E}_{FL}<10^{32}$
erg, any neutrino muon spectra where
$\bar{E}_{\bar{\nu}_{\mu}}<$100 MeV, or any proton-magnetic field
interaction may suppress somehow our estimates.  Therefore our
considerations are only preliminary and they must be taken with
caution  (also in view of the delicate chain of assumptions and
simplification).  We hope to stimulate with our work correlated
research in gamma-optical-neutron rays observations and neutrino
underground detectors in view of additional  solar activity. In
particular we suggest to control the very Super-Kamiokande data
records on $October- November$ solar flare X-radio peak activity,
namely on $26-28-30th$ October and $2nd-4th$ and $13$ November
X-ray onset (see figures below for time details). We like to
point  the attention to the hard X on set at $19:48$ U.T. on
$4th$ November $2003$.
\begin{acknowledgements}
The author wishes to thank Prof. M. Parisi,  M. Gasped,  and P.De
Sanctis Lucentin and M. De Santis and Drs. Cristina Leto  for
valuable suggestions.
\end{acknowledgements}
\begin{figure}
 \centering
   \includegraphics[width=\textwidth]{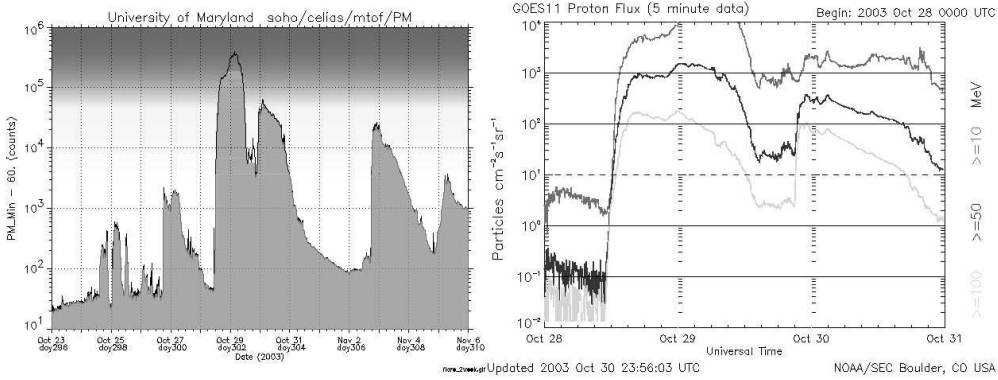}
   \caption{Proton solar flare flux (left: all the events during 23 October-6 November 2003 at lowest energies);
    (right:  detail of part (10-100 Mev) energy the spectra);
  the data are respectively from SOHO and GOES11 satellite experiments }
   \label{Fig:demo2}
\end{figure}

\begin{figure}
 \centering
   \includegraphics[width=0.4\textwidth]{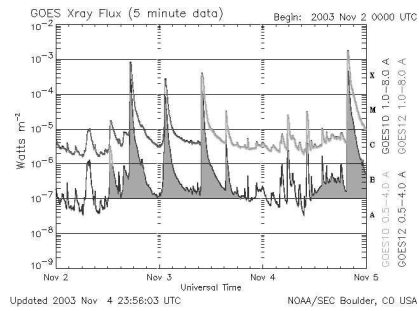}
   \caption{Left:X Solar Flare on-set 2nd-4th November 2003 by GOES satellite: readable are the X-peak outburst }
   \label{Fig:demo3}
\end{figure}

\begin{figure}
 \centering
   \includegraphics[width=0.5\textwidth]{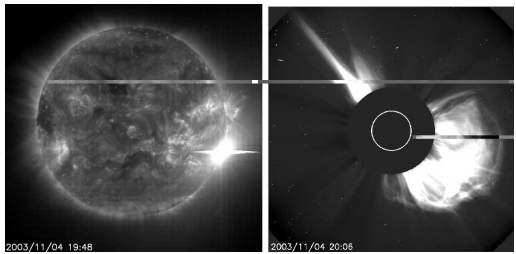}
   \caption{X Solar Flare on 4th November 2003 observed by GOES and SOHO satellites: note the  sharp outburst at 19:48-20:08 UT }
   \label{Fig:demo4}
\end{figure}



\label{lastpage}

\begin{thebibliography}{10}
\bibitem{ref0}{ F.Bachelet and A.M. Conforto, Nuovo
Cimento,3.1153(1956);  J.Simpson, Proc.National Acad. of Sc.of
USA,Vol43,42,(1957).}
\bibitem{ref1}{M. Alessio,L. Allegri, D. Fargion, S. Improta, N. Iucci, M. Parisi, G.
Villoresi,N.L.Zangrilli, Il Nuovo Cimento, 14C, 53-60, (1991).}
\bibitem{ref2}{M. Dryer, Space Science Reviews, 15 (1974), 403-468.}
\bibitem{ref3}{V.S. Berezinsky, C. Castagnoli and P. Galeotti,Il Nuovo Cimento, 8C, 185 (1985).}%
\bibitem{ref4}{D. Fargion, Maxim Khlopov; Astroparticle Vol.19, 3,p.441-446,(2003)}
\bibitem{ref5}{M.E. Machado and J. L. Linsky, Solar Phys., 42, 395 (1975).}%
\bibitem{ref6}{A. Dar and S. P. Rosen, Preprint 27803 - Los Alamos Th. Div., August (1984).}%
\bibitem{ref7}{T. K. Gaisser, T. Stanev, G. Barr, Preprint Bartol Research Inst., 22/01/89, BA-88-1.}
\bibitem{ref8}{W.C. Haxton, Phys. Rev. D, 36, 2283, (1987).}
\bibitem{ref9}{S.N. Karpov,L.I.Miroshnichenko,E.V.Vashenyuk, Il Nuovo Cimento, 21C, 551, (1998)}
\bibitem{ref10}{D. Fargion; adsabs.harvard.edu/abs/1989STIN...9023331F; Preprint INFN n.$721$; $19Dec$.(1989)}
\bibitem{Fargion 2002}{D. Fargion , Ap. J. 570, 909-925;(2002) }
\bibitem{bemporad} {C.~Bemporad, G.~Gratta and P.~Vogel, Rev.\ Mod.\ Phys.\  {74}
297(2002)}
\bibitem{strumia} {Strumia A., Vissani F., astro-ph/0302055, (2003)}
\bibitem{anti-matter}{D. Fargion and M.Khlopov; Astroparticle  Vol.19, 3,p.441-446,(2003)}.
 \bibitem{89} {L.I.Miroshnichenko, C.A.De Koning,R.
 Perez-Enriquez;SpaceScienceReviews 91; 615-715,(2000)}
 \bibitem{Lin} {R. P. Lin, et all.Ap.J, Volume 595, Number 2, Part 2;(2003)}.
 \bibitem{Bodek} {A.Bodek, H.Budd and J.Arrington hep-ex/0309024,
 (2003)}

\end{thebibliography}
\end{document}